# Experimental implementation of time-coding quantum key distribution

*William Boucher and Thierry Debuisschert*

THALES Research and Technology – France

Route Départementale 128

91767 Palaiseau Cedex, France

thierry.debuisschert@thalesgroup.com

**Abstract** : We have implemented an experimental set-up in order to demonstrate the feasibility of time-coding protocols for quantum key distribution. Alice produces coherent 20 ns faint pulses of light at 853 nm. They are sent to Bob with delay 0 ns (encoding bit 0) or 10 ns (encoding bit 1). Bob directs at random the received pulses to two different arms. In the first one, a 300 ps resolution Si photon-counter allows Bob to precisely measure the detection times of each photon in order to establish the key. Comparing them with the emission times of the pulses sent by Alice allows to evaluate the quantum bit error rate (QBER). The minimum obtained QBER is 1.62 %. The possible loss of coherence in the set-up can be exploited by Eve to eavesdrop the line. Therefore, the second arm of Bob set-up is a Mach-Zender interferometer with a 10 ns propagation delay between the two path. Contrast measurement of the output beams allows to measure the autocorrelation function of the received pulses that characterizes their average coherence. In the case of an ideal set-up, the value expected with the pulses sent by Alice is 0.576. The experimental value of the pulses autocorrelation function is found to be 0.541. Knowing the resulting loss of coherence and the measured QBER, one can evaluate the mutual information between Alice and Eve and the mutual information between Alice and Bob, in the case of intercept-resend attacks and in the case of attacks with intrication. With our values, Bob has an advantage on Eve of 0.43 bit per pulse. The maximum possible QBER corresponding to equal informations for Bob and Eve is 5.8 %. With the usual attenuation of fibres at 850 nm, it shows that secure key distribution is possible up to a distance of 2.75 km, which is sufficient for local links.

## I. INTRODUCTION

Quantum key distribution (QKD) exploits the fundamental principles of quantum mechanics to distribute securely a cryptographic key between two parties usually called Alice and Bob. It is an alternative method to algorithmic cryptography. The purpose of QKD is not to prevent a third party Eve to eavesdrop the line, but to make the eavesdropping detectable by Alice and Bob. In that case, they do not validate the key. The first protocols that have been proposed used the polarization basis of photons to encode the key, with either four non orthogonal polarization states [1] (usually called BB84) or two non orthogonal polarization states [2] (usually called B92). Since the first experimental demonstration [3], a lot of experimental demonstrations have been achieved based on those protocols [4]. Telecommunication applications of QKD require a propagation in an optical fiber. Technical limitations of optical fibers [5] makes polarization unpractical, and phase coding is preferred [6]. The principle is to implement a long arm Mach-Zender interferometer between Alice and Bob allowing each of them to modify the dephasing between the two arms of the interferometers. This technique allows a coding similar to that of BB84 with polarization. It is also necessary to compensate for polarization modifications due to the propagation but this can be achieved with go and return techniques that allow for birefringence compensation [7], [8].

In view of practical applications we have proposed a simple protocol based on time coding [9]. That method is attractive because it allows a simple implementation based on state-of-the-art optical components. Anyhow, that method as been few investigated. A related protocol based on pulses time-shifts has been proposed recently [10]. The importance of using spectral bandwidth limited pulses has been emphasized subsequently [11]. Another recently proposed protocol makes use of phase coherent pulses produced with a mode-locked laser [12].

Our protocols make use of coherent one photon pulses with square profile and duration T. Coherent means that the pulses can be modeled with pure states in the sense of quantum mechanics. As a consequence, they have a minimum time-frequency uncertainty product upon which the security relies. In the simplest protocol, two kinds of pulses (b and c) are sent by Alice with a delay of 0 (bit 0) or T/2 (bit 1) with respect to a clock [9] (FIG. 1). The detection by Bob may occur in three successive time slots (3, 4 and 5) of duration T/2. An eavesdropper may induce errors if he resends a pulse of duration T after detection in time slot 4, thus leading to an ambiguous result. In order to detect a possible shortening of the pulses resent by Eve, half of the pulses received by Bob are sent at random to a Mach-Zender interferometer having a path difference of cT/2. A measurement of the pulses coherence length allows measuring the average pulse duration [9]. The other half of the pulses received by Bob are sent to a photon-counter that measures precisely the photon time detection in order to establish the key. A four states protocol, with two additional pulses (a and d) carrying no information, imposes additional symmetry constraints to Eve, which keeps her to exploit the possible losses of the quantum channel [9].

From an experimental point of view, this protocol has several advantages. Available photon-counters can have response time smaller than one ns [13]. We will thus consider pulse durations in the 10-20 ns range for which the time propagation of the pulses is only little affected by the propagation disturbances of the fiber. A low error rate requires precisions in the arrival time of about 1 ns which makes it insensitive to fiber thermal dilatation. In addition pulse spreading due to group velocity dispersion starts to be noticeable only in the ps range with usual telecommunication fibers. The measurement of the arrival time of the photon does not require that the polarization of the photon is conserved. If the interferometer is made insensitive to the polarization, the whole system is potentially insensitive to the polarization. As a consequence there is no need for go and return of the pulses, which opens the way to high transmission rates. Coherent faint pulses can be produced combining a single mode diode laser and an high-speed electro-optics amplitude modulator which can be driven with an electrical pulse generator having rise time and decay time smaller than 1 ns. In addition, we propose a statistical method to measure the contrast of the interferometer which does not require that the phase be stabilized. These technical considerations combined with the advantages of the principle described above make the time coding protocol a realistic method for quantum key distribution.

In the following, we analyze the experimental requirements in order to implement our protocol, and we describe an experimental realization. We then use our experimental results to analyze the security in the case of intercept-resend attacks and in the case of attacks with intrication. We show that secure key distribution is possible.

## II. PROTOCOL

The security of the protocol can be quantified evaluating the mutual informations between Alice and Eve ($I_{AE}$) and between Alice and Bob ($I_{AB}$). The criterion $I_{AB} \geq I_{AE}$ [14] allows determining the parameters values for which the security can be guaranteed. $I_{AB}$ is a decreasing function of the quantum bit error rate (QBER denoted Q) calculated on the sifted key [9]. One specificity of our protocols is that Eve can exploit a possible lack of coherence of the pulses sent by Alice or the imperfections of Bob interferometer to get some information on the key. We call $g(t)$ the normalized temporal profile of the pulses that are launched by Alice. The autocorrelation function of $g(t)$ for a time delay $\tau$ is defined by:

$$\gamma(\tau) = \int dt\, g(t) g(t-\tau)$$

(1)

We denote $\gamma_{th}$ the value calculated for a delay corresponding to the propagation time difference of the two arms of the interferometer (in our case 10 ns). $\gamma_{th}$ quantifies the maximum coherence that can be measured. The value of $\gamma_{exp}$ that can be obtained from measurements may be smaller than $\gamma_{th}$. We define a parameter $\Delta$ that characterizes the relative loss of coherence due to the experimental imperfections. $\Delta$ (corresponding to $dC$ in [9]) is defined by:

$$\Delta = \frac{(\gamma_{th} - \gamma_{exp})}{\gamma_{th}}$$

(2)

Therefore, $I_{AE}$ is an increasing function of Q depending on the parameter $\Delta$ [9].

Once the parameters governing the mutual informations have been defined, one has to study the attacks that can be used by Eve to get some information on the key. We will first evaluate the security for the two kinds of intercept resend attacks we have studied in ref. [9]. Then we will propose an improvement of the protocol and study more sophisticated attacks where Eve intricates her state with the one transmitted from Alice to Bob. That leads to more efficient attacks that are expected to be close to optimal individual attacks.

Considering first the intercept-resend attacks, we suppose Eve has detected a photon in time-slot j. In the first attack, called two time-slots attack, Eve resends a pulse spanning two adjacent time slots j, j+1 or j, j-1. In the second one, called maximum coherence attack, Eve resends a pulse spanning on time-slots j-1, j, j+1. This way, she maximizes the coherence measured by Bob. This attack allows Eve to extract more information without being detected. The detection of that attack would require an additional interferometer with c.T path difference that is not implemented in our set-up. Therefore we have to consider both attacks. Simultaneous measurement of the QBER on the sifted key and of the relative loss of coherence $\Delta$ allows to evaluate the security of the transmission for a given kind of attack.

An simple approach of those attacks can be given. We focus on the maximum coherence attack which appears to be the most efficient of both attacks. Alice sends at random with equal probabilities the states corresponding to bit 0 and bit 1. With the notation of FIG. 1, they can be written $|bit\,0\rangle = \frac{\sqrt{2}}{2}(|3\rangle+|4\rangle)$ and $|bit\,1\rangle = \frac{\sqrt{2}}{2}(|4\rangle+|5\rangle)$. When Eve intercepts in time-slot 3 or time slot 5 she knows exactly which bit was sent by Alice. If she detects in time-slot 4, she does not know which bit was sent. When Eve intercepts in time-slot j (j being equal to 3, 4 or 5), she sends the state $|\psi_j\rangle$ defined by :

$$|\psi_j\rangle = \sqrt{\frac{x}{2}}|j-1\rangle + \sqrt{1-x}|j\rangle + \sqrt{\frac{x}{2}}|j+1\rangle$$

( 3 )

The mutual information between Alice and Eve is the ratio between the probability that Eve intercepts in time slot 3 or time slot 5 and Bob validates his measurement over the total probability that Bob validates his measurement. From the probability for Eve to detect in time-slot 3 or time-slot 5 and the expression of $|\psi_j\rangle$, one finds that the first probability is $\frac{1}{2}(1-x)$. To obtain the second probability one has to add to the first one the probability that Eve detects in time-slot 4 and Bob validates his measurement. That probability is equal to $\frac{1}{2}x$. The resulting total mutual information between Alice and Eve is thus $(1-x)$.

The quantum bit error rate is defined as the ratio of the error probability to the total probability that Bob validates his measure. This situation occurs when Alice sends, for example, a bit 0, Eve detects in time-slot 4 with probability $\frac{1}{2}$ and resends a state $|\psi_4\rangle$. The error probability is the probability for Bob to detect in time-slot 5 which is equal to $\frac{1}{4}x$. The total probability for Bob to validate his measure being $\frac{1}{2}$, the quantum bit error rate is $\frac{1}{2}x$.

The coherence of the pulse resent by Eve is measured with a Mach-Zender interferometer where the time propagation between the two arms is T/2. This results in superposing the state $|\psi_j\rangle$ and the same state where j is replaced by j+1. Integrating over all possible time slots, the coherence that is measured is $\sqrt{2(1-x)x}$. In the intercept-resend attack, we have considered that Eve intercepts with a probability m and resends a pulse each time she intercepts. Therefore, the QBER, mutual information and contrast are given by the following expressions :

$$QBER = Q = \frac{1}{2}mx$$

(4)

$$I_{AE} = m(1-x)$$

(5)

$$C = m\sqrt{2(1-x)x} + \frac{1}{2}(1-m) = \frac{1}{2}(1-\Delta)$$

(6)

The expression of the contrast features the parameter $\Delta$ that characterizes the relative loss of coherence and that has been defined previously. The parameter m being smaller than 1, the following relation has to be fulfilled :

$$I_{AE} + 2Q = m \leq 1$$

(7)

The parameter m can then be eliminated from the equations in order to obtain the expression of $I_{AE}$ as a function of $Q$ and $\Delta$ :

$$I_{AE} = 6Q + \Delta + 4\sqrt{Q(2Q+\Delta)}$$

(8)

The corresponding curves are represented on FIG. 5 for three experimental values of $\Delta$ (curves $a_3$, $b_3$ and $c_3$). They show that the protocol can stand intercept-resend attacks under experimental conditions.

Before considering more elaborated attacks implying an intrication between the state of Eve and the state transmitted from Alice to Bob, one must point out that the protocol can be improved so that the contrast value expected by Bob is 1 in the ideal case. This is obtained by measuring the arrival time of each photon in the output ports of the interferometer instead of integrating all of them and to selecting only those that actually allow to measure the coherence of the received pulses. For that we add an additional step in the protocol following the step where Bob has indicated to Alice all the measurements he has validated (corresponding of measurements in time slots 3 and 5). Bob indicates all pulses participating to the contrast measurement (and which, therefore, do not participate to the key). Then Alice reveals the corresponding bits. When Bob knows that a bit 0 has been sent, he only keeps the result if a photon has been detected in the time slot 4 in one of the two output ports. Otherwise he rejects the measurement result. In the case when Alice has sent a bit 1, he only keeps the results corresponding to a detection in time-slot 5. Doing that, the expected contrast averaged over all contrast measurement results is 1 in the ideal case. With that improved protocol, the coherence that is measured when Eve intercepts the pulses is now equal to $\frac{2\sqrt{2(1-x)x}}{2-x}$. In the particular case where no contrast loss is allowed ($\Delta=0$), the coherence has to be equal to one, which imposes x to be equal to $\frac{2}{3}$. Reporting that value in the expression in $|\psi_j\rangle$, one obtains that the pulses resent by Eve must have a uniform amplitude over the time-slots j-1, j, j+1, corresponding to a detection probability of $\frac{1}{3}$ in one of those time-slots. Reporting that value of x in the expressions of Q an $I_{AE}$ and eliminating m, one finds that the mutual information between Alice and Eve is now given by :

$$I_{AE} = Q$$

( 9 )

That expression shows that the new step we have added to the protocol actually improves security in the case of intercept-resend attacks by diminishing the information available to Eve.

More general attacks can be considered that may allow Eve to extract more information about the key. The basic idea is to consider the most general unitary transform that can be performed by Eve and that implies that she intricates her state with the one transmitted from Alice to Bob. Doing that, optimal individual attacks have been studied in the case of the BB84 protocol (see ref. [4] for a review) or in the case of the symmetric 3 states protocols [15]. In order to give a general evaluation of the security of our system, we have performed a preliminary analysis of those attacks for our protocol. We expect those attacks to be close to the individual optimal attacks. The purpose of the paper is mainly to describe the experimental set-up and to describe the exploitation of its results. Therefore, we will not give here the detailed derivation of such attacks but only sketch the main lines of the derivation. The bits sent by Alice are described in a 3 dimensions Hilbert space. In order to maximize the coherence, Eve transforms the state $|j\rangle$ in a superposition of states $|j-1\rangle$, $|j\rangle$ and $|j+1\rangle$. The most general unitary transform describing the intrication of Eve with Bob thus requires that Eve's Hilbert space has 9 dimensions. The state sent by Eve has to fulfil the constraints imposed by the contrast measurement at Bob. With the new step added to our protocol, the contrast can be equal to 1 in the ideal case. Therefore, Eve has to optimize the state so that its coherence be maximized. That constraint results in decreasing the dimension of Eve Hilbert space, in which all states can be calculated. The contrast can then be optimized by adjusting the remaining parameters. As a result, it is possible to calculate the maximal information available to Eve as a function of the QBER supposing that Eve interacts individually with all the states launched by Alice. The resulting curves are displayed on FIG. 6 for the values of $\Delta$ resulting from our experiment. They show that attacks involving intrication (solid curves) are more efficient than intercept-resend attacks (dashed curves). In the case where no contrast loss is allowed (curve a, $\Delta=0$), the maximum QBER value is equal to 12 %, what is similar to the limit of the BB84 protocol. In the case of an imperfect set-up with a possible contrast loss, the mutual information between Alice and Eve increases rapidly (curves b and c), showing the importance of a good coherence and a good contrast measurement in the experiment. Anyhow, even in the case of an imperfect set-up, secure key transmission is still possible with our improved protocol.

### III. EXPERIMENTAL SET-UP

The purpose of the study is to evaluate the possible experimental implementation of such protocols and to deduce if secure key transmission is possible. Several practical problems have to be solved in order to implement a realistic set-up. Alice has to produce pulses being as close as possible to the Fourier transform limit, taking into account the time responses of the driving electronics and of the photon detectors as well as the frequency linewidth of conventional laser diodes. The pulses must be long enough not to be affected by the group velocity dispersion of conventional optical fibers. The detection time of the photon has to be precisely measured and synchronized with the reference time of Alice. An interferometer has to be built in order to precisely measure the average autocorrelation of the received pulses. In a practical system, the source at Alice and the interferometer at Bob will be distant and, moreover, Bob will receive only faint pulses from Alice. Therefore, active stabilization of the phase of the interferometer is unrealistic and a method to measure $\gamma$ that can accommodate a phase drift has to be found. In addition, one must take into account that the sequences of pulses are limited in time, thus the number of events is limited. In the following, solutions to those practical problems are proposed in order to build an experimental set-up allowing secure transmission.

The experimental set-up is depicted on FIG. 2. The operating wavelength is chosen mainly for the availability of Si photon-counters (EG&G SPCM 14) having a high quantum efficiency (50 %), a short rise-time (300 ps) and 100 dark-counts per second in the near infrared. Therefore, the source we use is a DBR diode laser (Spectra Diode Labs SDL 5702) producing a cw beam at 852 nm. The diode is mounted with an external grating (1200 grooves/mm) in Littman configuration in order to improve the frequency linewidth and the optical isolation. The specified instantaneous frequency linewidth is 3 MHz. The measured frequency drift is 10 MHz over typical time durations of 10 ms. For a 3 m path difference interferometer, the corresponding phase shift during the launch of a sequence of pulses (3.2 ms) is 0.2 rad. It is small enough to consider that the phase of the interferometer is constant during one sequence. After passing through a Faraday isolator, the beam is coupled into a single-mode optical fiber. An integrated-optics amplitude modulator (UTP) with 8 GHz bandwidth and 30 dB extinction ratio allows tailoring an optical pulse with desired duration and delay. An electronic generator (Lecroy LW 410 A) that can produce pulses with rise-time and decay-time of 3 ns drives it. Therefore the pulses duration T is chosen around 20 ns in order to have optical pulses with a profile as square as possible and a duration much longer than the time-response of the photon-counters. As a consequence, the time-slot duration (T/2) is chosen equal to 10 ns as well as the propagation time difference between the two arms of the interferometer (corresponding to a 3 m path difference). The frequency linewidth corresponding to the 20 ns time duration of the pulses is 50 MHz, which can be considered large compared to the specified 3 MHz instantaneous frequency linewidth of the diode laser. A free-space optical attenuator allows tuning the average number of photon per pulse to the required value. The optical pulses are then propagated into a single-mode optical fiber and sent to Bob.

At the entrance of Bob, the beam is collimated with a beam waist of 2 mm in order to have a Rayleigh length of 14.8 m much larger than the interferometer path difference. From then on the set-up is free-space. A polarizer ensures that the beam entering the interferometer has a constant polarization. Bob set-up mainly consists of two arms where the pulses are sent at random thanks to a 50/50 beamsplitter. The first one comprises a photon-counter in order to measure the detection time of the photons. The second one is a Mach-Zender interferometer with a 3 m path difference allowing measuring the average autocorrelation of the pulses thanks to a contrast measurement. In the photon counting regime, the contrast of the interferometer is measured by the difference between the output photon numbers normalized to their sum. Two photon-counters with balanced efficiencies measure the number of photons in each output port. During one sequence of pulses (3.2 ms), the total number in each output port is measured with an electronic counter (Stanford Research SR400). The Si Photon-counters that are used to measure the photons time-detections and the numbers of photons in the outputs of the Mach-Zender interferometer are protected from parasitic light with interferential filters having a transmission of 50 % at their central frequency 850 nm and a band-pass of 10 nm. The photon-counters dead time after a triggering is 50 ns. Therefore the period between the launch of two successive pulses has been chosen equal to 100 ns. During the launch of the pulses, the chronogram of the photon detections is registered with a 8 Msample memory depth digitizing oscilloscope and stored in its memory. A common external generator synchronizes Alice's pulse generator and Bob's oscilloscope. The resulting file is then transferred to a computer for post-processing. That method allows to keep a precise timing for all successive pulses. The oscilloscope resolution has been chosen equal to 400 ps close to that of the photon-counters. Therefore pulses can be sent in sequences with a maximum duration of 3.2 ms due to the 8 Msample of the memory depth. In addition, due to delays internal to the electronic counter, the delay between two successive sequences is equal to 5 ms.

# IV. SET-UP CHARACTERIZATION

## A. Intense pulses

The set-up is first characterized using intense pulses in order to measure the actual pulse profile and to characterize the interferometer visibility. The measured pulse intensity profile (FIG. 3) can be fitted with a hyper-Gaussian function plus a background accounting for the imperfect extinction of the modulator. That function has the following expression :

$$I(t) = I_A \exp\left(-\frac{t^{2n}}{2\sigma^{2n}}\right) + I_B$$

( 10)

The ratio between $I_B$ and $I_A$ is 1e-3, corresponding to the 30 dB extinction ratio of the modulator. The experimental pulses have a FWHM of 18.7 ns corresponding to $\sigma$ equal to 9.6 ns. The order of the hyper-Gaussian is n=4. We can then calculate the normalized pulse amplitude $g(t)$ which is proportional to $\sqrt{I(t)}$ and normalized over a domain of 100 ns corresponding to the delay between two consecutive pulses. This function allows calculating the value of $\gamma_{th}$ in the case of a perfect experiment. Due to the background and the wings of the pulses actual profile characterized above, the expected value of $\gamma_{th}$ is 0.576 instead of 0.50 for perfect 20 ns squares. The fringe visibility of the interferometer measured with intense pulses leads to a value of $\gamma$ of 0.54. The difference with the theoretical value is due to imperfections such as lack of coherence of the source, mirrors defects or imperfect mode matching of the beams. As a result, the relative loss of coherence $\Delta_{int}$, evaluated with intense pulses, is equal to 0.061.

The propagation delay between Alice and Bob is roughly measured to be 120 ns sending a 400 µs long sequence of 10000 intense pulses that can be detected with a standard high-speed photodiode. The directly measured pulses arrival times are typically distributed over 20 ns. It shows that the pulse generator and oscilloscope 10 MHz internal clocks have a relative frequency difference of 5e-5. The width of that distribution can be reduced to 400 ps (oscilloscope resolution) with a fine adjustment of the pulses period involved in the post processing. Doing that, one can obtain a precise synchronization between Alice and Bob.

## B. Photon counting regime

After characterizations with intense pulses, one has to characterize the set-up in photon counting regime. In order to measure the experimental quantum bit error rate (QBER), Alice sends 3.2 ms long sequences of pulses with alternate delays 0 ns (bit 0) and 10 ns (bit 1). The attenuator is tuned in order to get 0.1 photon per pulse in average at the output of Alice. The error is the ratio between the number of detections in the wrong time-slot (e.g time-slot 5 for bit 0) to the total number of non ambiguous detections (time-slots 3 and 5). The QBER value is found by post processing the data finding the minimal error value as a function of the propagation delay between Alice and Bob. Furthermore, it can be optimized as a function of the pulse width in order to reduce the influence of the wings of the pulses. A FWHM of 18.7 ns has been chosen since it leads to an expected QBER value in the 2.2 % range and it does not depart too much from the 20 ns value for ideal pulses.

The second parameter that has to be evaluated is the average autocorrelation function $\gamma_{\exp}$ in photon counting regime. This is done with a contrast measurement which is a macroscopic parameter measured over numerous pulses. One must point out that the pulses that are used in the contrast measurement do not contribute directly to the secret key. Therefore any imperfection in the contrast measurement will not compromise the secret key, but will be taken into account via the value of $\Delta$. $\gamma_{\exp}$ takes into account all the experimental imperfections. The expected contrast is given by $\gamma_{\exp}\cos(\phi)$, where $\phi$ is the phase difference between the two arms of the interferometer depending on the optical carrier frequency and of the time propagation difference between the two arms. We have seen previously that the phase drift is small during one sequence. Therefore we have assumed a constant phase when evaluating the contrast for a given sequence. Doing that we slightly underestimate the value of the autocorrelation function and we slightly overestimate the value of $\Delta$. As a consequence, we slightly overestimate the value of $I_{AE}$, which does not compromise the security of the key. On the opposite, the phase drift is large during two successive sequences. We can thus consider the phase as random for different sequences, the phase distribution being uniform for a sufficiently large number of sequences. The phase is thus let as a free-running parameter. We then exploit that property to calculate the pulse autocorrelation with a statistical method. As a consequence, it is not necessary to know precisely the phase of the interferometer and to stabilize it. Therefore, there is no control-loop between Alice and Bob, which results in a great simplification of the experimental set-up.

Moreover, in a given sequence, the output photon number is limited. Therefore, the contrast measurement is affected by shot-noise. In order to improve the accuracy in the evaluation of $\gamma_{\exp}$, measurements over $N_s$ successive sequences are performed. The output state of the Mach-Zender interferometer is calculated assuming a weak coherent state at the input [16]. Due to the shot-noise, the contrast can take values characterized by a gaussian probability density. Assuming that the average number of photons in each sequence $N_p$ is sufficiently large and the contrast mean value is not too close to 1, the variance of the contrast probability distribution has the expression $\sigma_k^2 = \left(1 - \gamma_{\exp}^2 \cos^2(\varphi_k)\right)/N_p$ where $\varphi_k$ is the phase of the $k^{\text{th}}$ sequence. For given values of $\gamma_{\exp}$ and $\varphi_k$, the probability density to measure a contrast value $C_k$ is thus :

$$p(C_k) = \frac{1}{\sqrt{2\pi}\sigma_k} \exp\left(-\frac{(C_k - \gamma_{\exp}\cos(\phi_k))^2}{2\sigma_k^2}\right)$$

From a given set of $N_s$ measured values of the contrast, one can deduce the probability density of $\gamma_{\exp}$ multiplying the probability density corresponding to each measurement, averaging over the phase distribution assumed uniform and applying Bayes theorem [17]. One obtains a gaussian centered on $\gamma_0$ given by $\gamma_0^2 = 2(\overline{C^2} - \sigma^2)$. $\overline{C^2}$ is the variance of the centered contrast measurements distribution. We have replaced $\sigma_k^2$ with $\sigma^2 = 1/N_p$ neglecting the dependence of $\sigma_k^2$ with $\gamma_{\exp}$, which results in a slight overestimation of the noise. That approximation is justified by the fact that $\gamma_{\exp}$ will be of the order of 0.5 in practical. The variance of the probability density of $\gamma_{\exp}$ is given by $\sigma_T^2 = 2/(N_p N_s)$. It shows that the total number of detected photons is taken into account resulting in an increase in the precision measurement of $\gamma_{\exp}$. The factor 2 in $\sigma_T^2$ results from the average on squared cosine of the random phases. One should point out that the statistical analysis of the contrast only results in a decrease of a factor $\sqrt{2}$ in the precision of its evaluation as compared to the case where the phase is stabilized and constant.

Then, knowing the distribution of $\gamma_{\exp}$, Bob can evaluate the minimum value of the pulses autocorrelation for a given level of security. A level of security meaning the probability to validate a key that has been eavesdropped without noticing it. For example, if he tolerates an error probability of 1e-3, the minimum value of $\gamma_{\exp}$ is not smaller than $\gamma_{3\sigma} = \gamma_0 - 3\sigma_T$. Replacing $\gamma_{\exp}$ with $\gamma_{3\sigma}$ in the expression of $\Delta$ allows to report the corresponding $I_{AE}$ curve on the security diagram (FIG. 5 and FIG. 6). Bob can then either calculate his information advantage on Eve for the measured QBER or evaluate the maximum QBER allowing security from the relation $I_{AB}=I_{AE}$. Before sending the key, Alice and Bob have to integrate enough sequences in order to reduce sufficiently the statistical error down to the value of $\Delta$ that would be obtained with intense pulses. Once the required level has been reached, Alice and Bob can start exchanging keys safely while keeping measuring the contrast.

## V. EXPERIMENTAL RESULTS

Applying that method, we have registered 290 successive sequences with an average of 282.5 photons per sequence at the output of the interferometer. That leads to a noise variance equal to $\sigma^2 = 3.54e-3$ and $\sigma_T^2 = 2.44e-5$. The difference with the actual number of photons launched by Alice is due to the beamsplitter, interferential filters transmission, photon counter efficiency and spurious loss in the set-up. The contrast measurements are reported on FIG. 4. The variance of the resulting centered contrast distribution is equal to $\overline{C^2} = 0.15$. The central value of the autocorrelation distribution is thus equal to $\gamma_0 = 0.541$. That value is the same as that measured with intense pulses. With our experimental data, $\sigma_T$ is equal to 4.9e-3, which leads to $\gamma_{3\sigma} = 0.526$.

From that value, one can calculate the coherence loss relative to the theoretical value $\Delta$. The theoretical value calculated from the actual shape of our pulses is $\gamma_{th} = 0.576$. With $\gamma_{exp}$ equal to $\gamma_{3\sigma}$, the resulting value of $\Delta$ is 0.086. Taking only into account the imperfections of the set-up or for an infinite number of photons, $\gamma_{exp}$ is equal to $\gamma_0$ and $\Delta$ has the value of 0.061 obtained with intense pulses. It shows that the statistical noise it not dominant as compared with the imperfections of the set-up. It could in addition be reduced up to being negligible increasing the number of sequences.

In our set-up, all photons detected in the output ports of the interferometer are integrated to evaluate the contrast. Taking into account that the statistical corrections to the evaluation of $\Delta$ are small and that the experimental imperfections are independent of the data processing, we have applied the experimental values of $\Delta$ in the case of the improved protocol where the photon detections in the output ports of the interferometer are selected. For both intercept-resend attacks and for the attack with intrication, the information advantage of Bob on Eve has been calculated with statistical noise taken into account (FIG. 5, curves $c_2$ and $c_3$, and FIG. 6, curve c, corresponding to $\Delta = 0.086$), with no statistical noise (FIG. 5, curve $b_2$ and $b_3$, and FIG. 6, curve b, corresponding to $\Delta = 0.061$) and in the case of a perfect system (FIG. 5, curve $a_2$ and $a_3$, and FIG. 6, curve a, corresponding to $\Delta = 0$). For each curve one can calculate the maximum acceptable QBER value given by the relation $I_{AB}=I_{AE}$. The results are collected in Table I.

In the same series of measurements as those made to evaluate the contrast, we have measured the QBER of the last sequence. We have found a value of 3.3 %. It is a little higher than the expected value of 2.2 %. It is attributed to a slightly imperfect tuning of the modulator extinction. The information advantage of Bob on Eve has been calculated for a QBER of 3.3 % for the attacks considered. The results are collected in Table II. In all cases Bob has more information than Eve on the key. In the less favorable case (taking into account statistical noise) this advantage is only 0.22 bit / pulse. We have found the same value for the maximum coherence attack and for the attack with intrication. The higher efficiency of the latter attack is compensated by the improvement of the protocol in that case.

In order to measure the minimum QBER available with our set-up, we have registered five successive sequences. The average total number of photons per sequence was 411, the useful average number of photons detected in time-slots 3 and 5 was 194 per sequence. The measured QBER was $1.62 \pm 0.75$ %. It shows clearly that a small QBER can be obtained even if the pulses do not have a perfect profile. We have evaluated the information advantage of Bob on Eve for that value. The results are gathered in Table III. In the less favorable case (attack with intrication taking into account statistical noise), this advantage is 0.43 bit / pulse. The gain of a factor 2 compared with the previous measurement shows the importance to optimize the set-up in order to have a QBER as low as possible. Inspecting the curves $b_3$ and $c_3$ on FIG. 5 or b and c on FIG. 6 shows that the statistical noise has few influence since the two curves are very close. There is few interest in integrating much more than what has been done unless a very low error probability is required. On the opposite, since the slopes of those curves are steep, it is much more efficient to decrease the QBER in order to make the amount of information available to Eve as small as possible.

We can analyze the main origin of the noise in our set-up. The dark-counts rate has been measured to be 110

per sec, leading to a probability of 1.1e-6 to occur during a useful time-slot (10 ns). We have measured a level of parasitic light of 1000 counts / sec due to the imperfection of the interferometric filters. This leads to a probability of 1e-5 to detect a parasitic photon. We have detected 194 useful photons in average during one sequence (3.2 ms). One pulse is sent every 100 ns and there are two useful time slots of 10 ns corresponding to each pulse. Therefore, the probability to detect a useful photon in one time-slot is 3e-3 . The integrated optics modulator has an extinction ratio of 30 dB, leading to a probability of 3e-6 to detect a background photon during one time slot. The probabilities corresponding to those three identified sources of noise are comparable within a factor of ten. In order to reach the limit imposed by the photon counter dark-counts, one should improve the elimination of parasitic light and the rejection of the background. Since it is difficult to improve the extinction ratio of the modulator, one can imagine a double pass configuration in order to reduce the background. Doing these improvements would lead to a noise dominated by the photon counter dark-counts. In the case of a propagation through a fiber, that noise will become more and more important as the propagation length increases and the line transmission decreases. That would result in an increase of the QBER reducing the security of the set-up [18]. We can roughly estimate the range of our set-up comparing our best measured QBER (1.62 %) with the maximum allowed QBER in the case of the improved protocol and the attack with intrication (5.8 % in the least favorable case). With a noise level being constant (parasitic light), this allows an attenuation of 3.6 (5.5 dB) in the line. Considering a usual attenuation of 2 dB/km for 850 nm single mode fiber leads to a range of 2.75 km, which is sufficient for local links. That shows that such a set-up is useful for practical applications.

## VI. CONCLUSION

We have implemented an experimental set-up in order to demonstrate the feasibility of our time-coding protocol. Alice produces coherent 20 ns faint pulses of light modulating the beam of a cw diode laser emitting at 853 nm. Sequences of 32000 pulses are sent to Bob with delay 0 ns (bit 0) or 10 ns (bit 1). This defines three 10 ns time-slots leading to non ambiguous detection results for the first one and the third one and to ambiguous result for the second one. Bob directs at random the received pulses either to a first arm used to establish the key or to a second arm used to measure the coherence length of the received pulses. In the first arm, a 300 ps resolution Si photon-counter allows to precisely measure the detection time of each photon. Bob can then position each detection event within the corresponding time-slot and compare the result to the pulse sent by Alice. He can then infer the QBER due to pulse profile imperfections, background light, parasitic light or detectors dark-count. The minimum obtained QBER is 1.62 %. The second arm of Bob set-up is a Mach-Zender interferometer. It has a long arm having a 3 m long path difference with the short arm resulting in a 10 ns propagation delay. It allows to characterize the coherence of the received pulses characterized by their autocorrelation function. The contrast measurement of the output beams is directly related to the autocorrelation function of the pulses and then to the average duration of the received pulses. That allows Bob to detect an attack where Eve detects the pulses sent by Alice and resends to Bob pulses with shorter duration. In order to work with faint pulses, two Si photon-counter are used to measure the output photon numbers in each arm of the interferometer. The difference of that numbers normalized to their sum is a measurement of the contrast. Integrating over a large number of sequences (typically 290) allows to average out the phase dependence of the contrast. Taking into account the finite photon number per sequence (typically 282), one can finally deduce a Gaussian probability density for the experimental pulse autocorrelation. The central value has been measured to be 0.541, and the root-mean square 4.9 e-3. The expected theoretical autocorrelation value resulting from the shape of the pulses sent by Alice is 0.576. From those values one can calculate the parameter $\Delta$ that characterizes the loss of coherence relative to the ideal value. We have considered the situation where we allow a probability of 1e-3 not to detect the eavesdropper, what leads to a value of $\Delta$ of 0.086. The corresponding mutual information between Alice and Eve can be evaluated for the given value of $\Delta$ as a function of Q. We have considered two kinds of intercept-resend attacks and an attack with intrication that we expect to be close to the optimal individual attack. The less favorable situation corresponds to the case where Eve resends pulses on three adjacent time-slots, thus maximizing the coherence. Comparing with the mutual information of Alice and Bob, in the case of a QBER of 1.62 %, Bob still has an advantage of 0.43 bit per pulse on Eve. That shows that secure key transmission is possible with such a set-up. The maximum allowed QBER is 5.8 %. Supposing a constant noise level in our set-up that is mainly due to parasitic light, we find that an attenuation of 3.6 is possible. That allows a propagation range of 2.75 km that is sufficient for local links. Decreasing the parasitic light level down to the dark-count level of our photon-counter would allow to increase furthermore that range.

In view of practical applications, a improvement of the experiment would consist in transposing the experiment at 1550 nm using telecom fibers. The typical loss of 0.2 dB/km would result in an increase in the secure transmission range. One drawback of that wavelength range is the performances of InGaAs photon counters that are not as good as the Si photon counters used at 850 nm. Their efficiency is smaller than 10 % and the typical dark-count rate is typically 1e5/s. Those parameter values lead to typical secure range of 20 km in the case of intercept-resend attacks [9]. Other typical improvements would consist in using shorter pulses with sharper rise-time and decay-time (typically smaller than 1 ns). The use of a polarization free interferometer would allow transmitting keys through standard depolarizing optical fibers. Possible improvements in the protocol should lead to a better efficiency in the transmission and a simplification of the set-up. Therefore transmitting cryptographic keys over distances of 20 km at Telecom wavelength is realistic with our time coding protocol.

## VII.  ACKNOWLEDGEMENT

The authors are grateful to Philippe Grangier and Ariel Levenson for helpful discussions and advices on the experiment.

**FIGURE CAPTIONS :**

FIG. 1 : Principle of the time coding protocols. In the two states protocol, only pulses b and c, corresponding to bit 0 and 1 respectively, are emitted by Alice. In the 4 states protocols, two additional pulse a and d carrying no information can be launched by Alice. The detection by Eve of a photon in time-slot 4 does not allow her to discriminate between bit 0 and bit 1. Therefore, she cannot avoid to induce errors when resending a pulse.

FIG. 2 : Experimental set-up. The beam produced by a diode laser at 852 nm is modulated by Alice to produce coherent 18.6 ns pulses (FWHM). After attenuation, the faint pulses are sent to Bob through a single mode fiber. They are directed by Bob either to a photon counter to establish the key or to a Mach-Zender interferometer to measure their average duration. The long arm of the interferometer is 3 m long resulting in a 10 ns delay between the two arms. The time resolution of the photon counters is 300 ps.

FIG. 3 : Temporal profile of the pulses sent by Alice. The FWHM is 18.7 ns. An hyper-Gaussian fit allows to calculate the pulse autocorrelation function, which gives the maximum coherence expected with an ideal set-up.

FIG. 4 : Contrast measurement for 290 successive sequences of pulses. 282.5 photons per second are detected in average. The phase is not controlled, therefore it acts as a random variable. A statistical treatment of the data allows to deduce the autocorrelation of the received pulses and to compare it with the expected value in the case of an ideal set-up. That measurement allows to detect a possible modification of the length of the pulses resent by Eve.

FIG. 5 : Mutual informations as a function of the QBER (Q) in the case of intercept-resend attacks. $I_{AB}$ decreases with Q and is independent of the coherence loss $\Delta$. $I_{AE}$ is calculated for the two time-slots pulses attack (index 2) or the maximum coherence attack (index 3). In each case, $I_{AE}$ is calculated for the ideal case and the two experimental cases corresponding to the relative coherence loss $\Delta$ : 0, 0.061 and 0.086 (a, b and c respectively). The dashed line gives the maximum information that can be obtained by Eve when she intercepts all pulses sent by Alice. For a given value of $\Delta$, $I_{AE}$ is an increasing value of Q. The vertical lines corresponds to the 1.62 % and 3.3 % experimentally measured QBER values. Secure key transmission is possible when $I_{AB}$ is greater than $I_{AE}$.

FIG. 6 : Mutual informations as a function of the QBER (Q) in the case of the attack with intrication (solid line). In each case, $I_{AE}$ is calculated for the ideal case and for the two experimental cases corresponding to the relative coherence loss $\Delta$ : 0, 0.061 and 0.086 (a, b and c respectively). They are compared with the intercept-resend attacks

calculated with the same parameter (dashed line) and considering that Eve intercepts a fraction of the pulses ranging from 0 to 1. The thin line gives the maximum information that can be obtained by Eve when she intercepts all pulses sent by Alice. For a given value of $\Delta$, $I_{AE}$ is an increasing value of Q. The vertical lines corresponds to the 1.62 % and 3.3 % experimentally measured QBER values. Secure key transmission is possible when $I_{AB}$ is greater than $I_{AE}$.

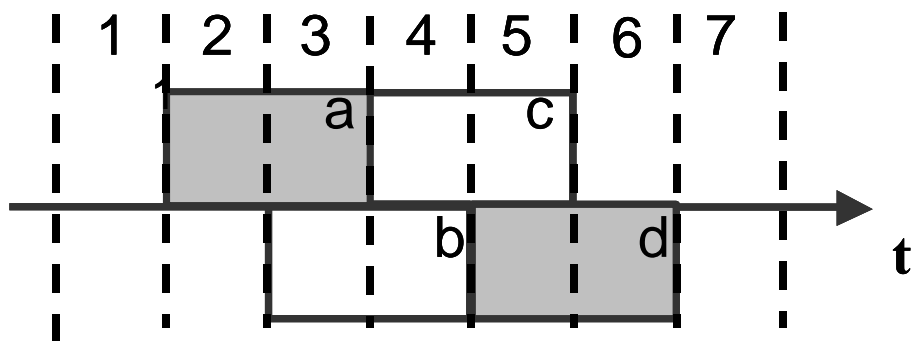

FIG. 1 : Principle of the time coding protocols. In the two states protocol, only pulses b and c, corresponding to bit 0 and 1 respectively, are emitted by Alice. In the 4 states protocols, two additional pulse a and d carrying no information can be launched by Alice. The detection by Eve of a photon in time-slot 4 does not allow her to discriminate between bit 0 and bit 1. Therefore, she cannot avoid to induce errors when resending a pulse.

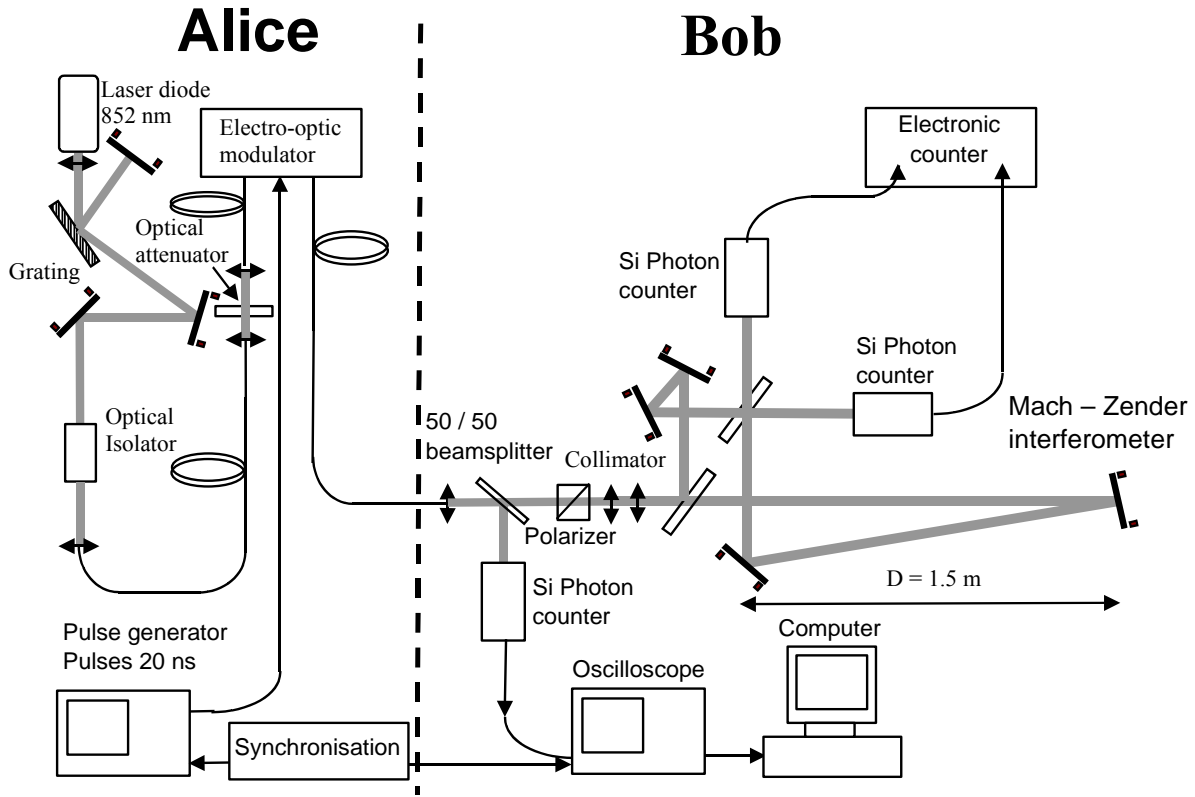

FIG. 2 : Experimental set-up. The beam produced by a diode laser at 852 nm is modulated by Alice to produce coherent 18.6 ns pulses (FWHM). After attenuation, the faint pulses are sent to Bob through a single mode fiber. They are directed by Bob either to a photon counter to establish the key or to a Mach-Zender interferometer to measure their average duration. The long arm of the interferometer is 3 m long resulting in a 10 ns delay between the two arms. The time resolution of the photon counters is 300 ps.

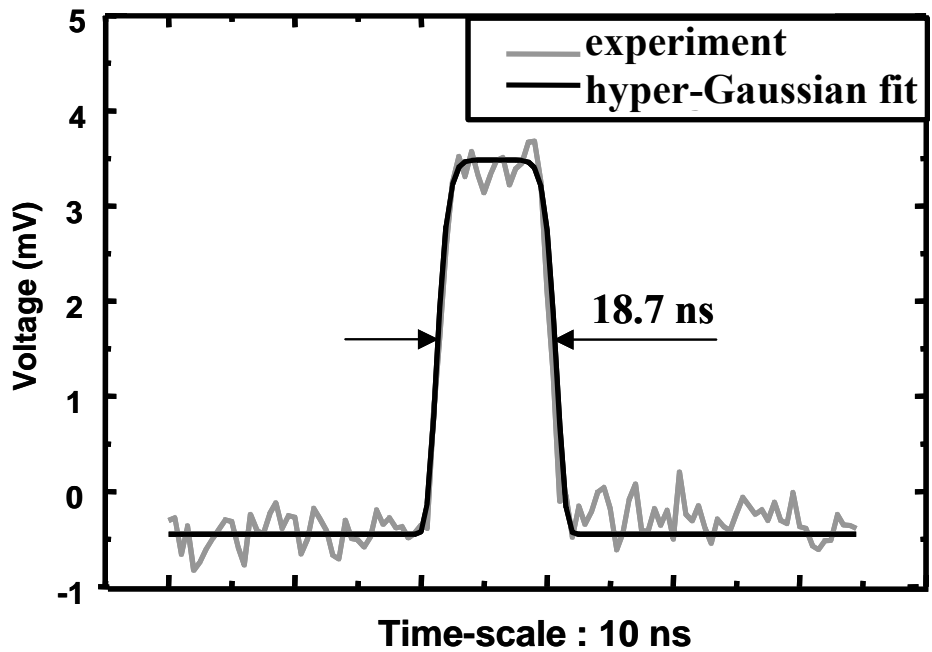

FIG. 3 : Temporal profile of the pulses sent by Alice. The FWHM is 18.7 ns. An hyper-Gaussian fit allows to calculate the pulse autocorrelation function, which gives the maximum coherence expected with an ideal set-up.

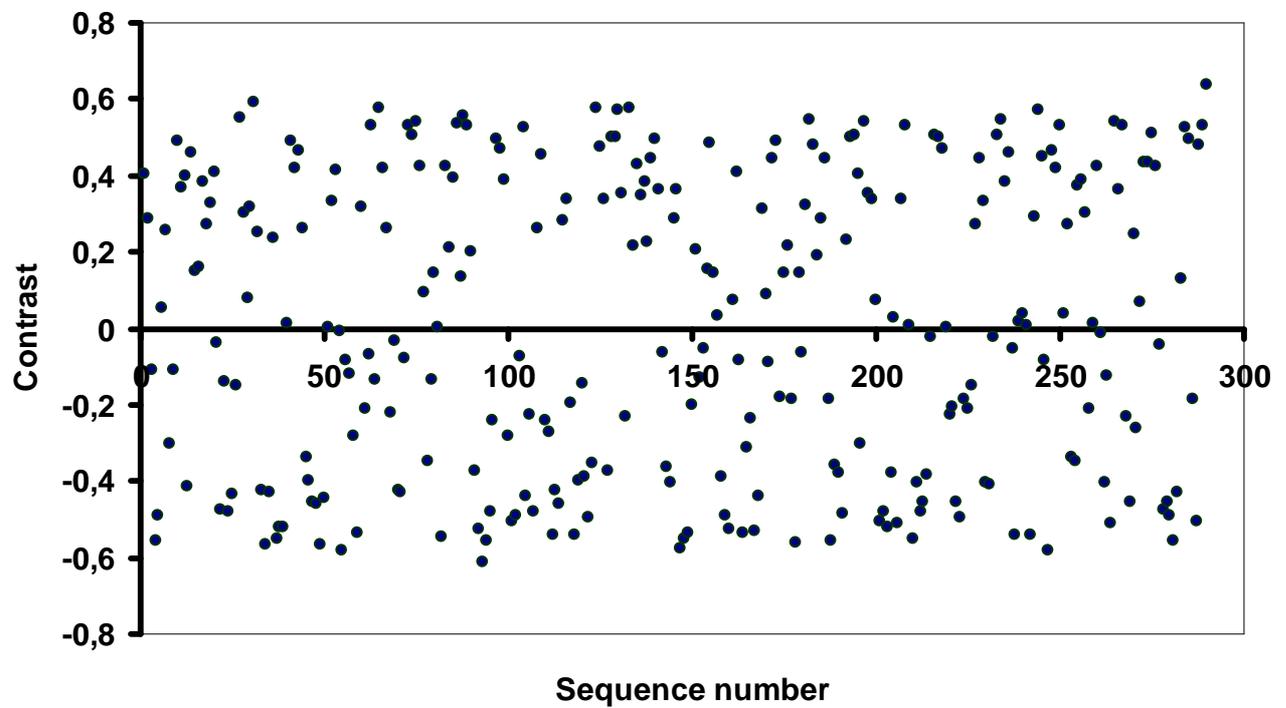

FIG. 4 : Contrast measurement for 290 successive sequences of pulses. 282.5 photons per second are detected in average. The phase is not controlled, therefore it acts as a random variable. A statistical treatment of the data allows to deduce the autocorrelation of the received pulses and to compare it with the expected value in the case of an ideal set-up. That measurement allows to detect a possible modification of the length of the pulses resent by Eve.

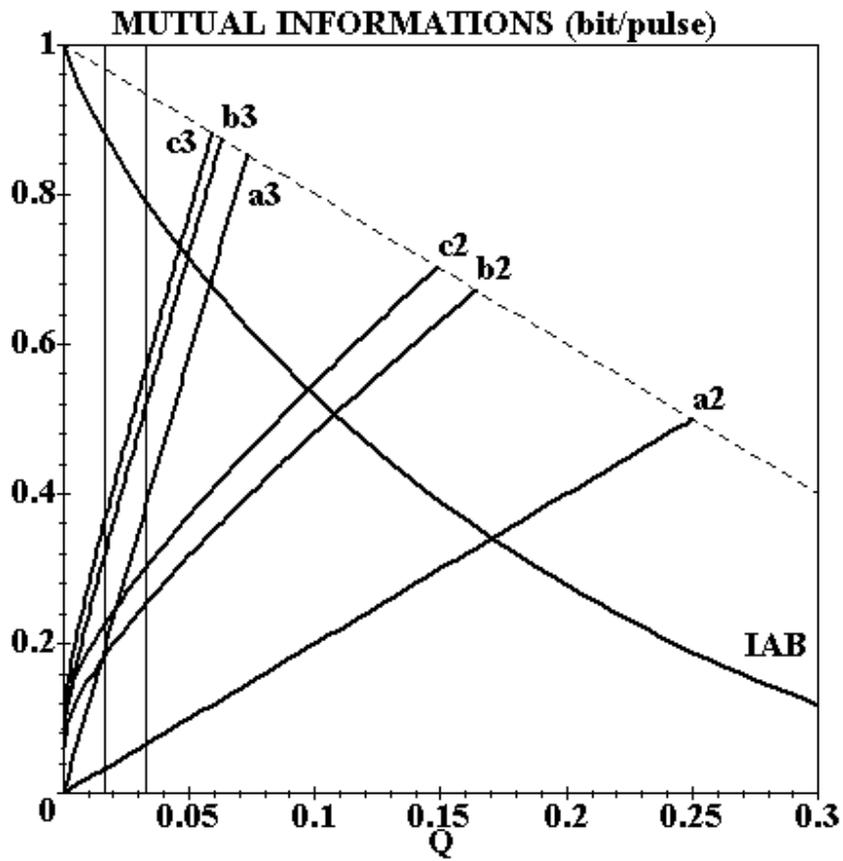

FIG. 5 : Mutual informations as a function of the QBER (Q) in the case of intercept-resend attacks. IAB decreases with Q and is independent of the coherence loss $\Delta$. IAE is calculated for the two time-slots pulses attack (index 2) or the maximum coherence attack (index 3). In each case, IAE is calculated for the ideal case and the two experimental cases corresponding to the relative coherence loss $\Delta$ : 0, 0.061 and 0.086 (a, b and c respectively). The dashed line gives the maximum information that can be obtained by Eve when she intercepts all pulses sent by Alice. For a given value of $\Delta$, IAE is an increasing value of Q. The vertical lines corresponds to the 1.62 % and 3.3 % experimentally measured QBER values. Secure key transmission is possible when IAB is greater than IAE.

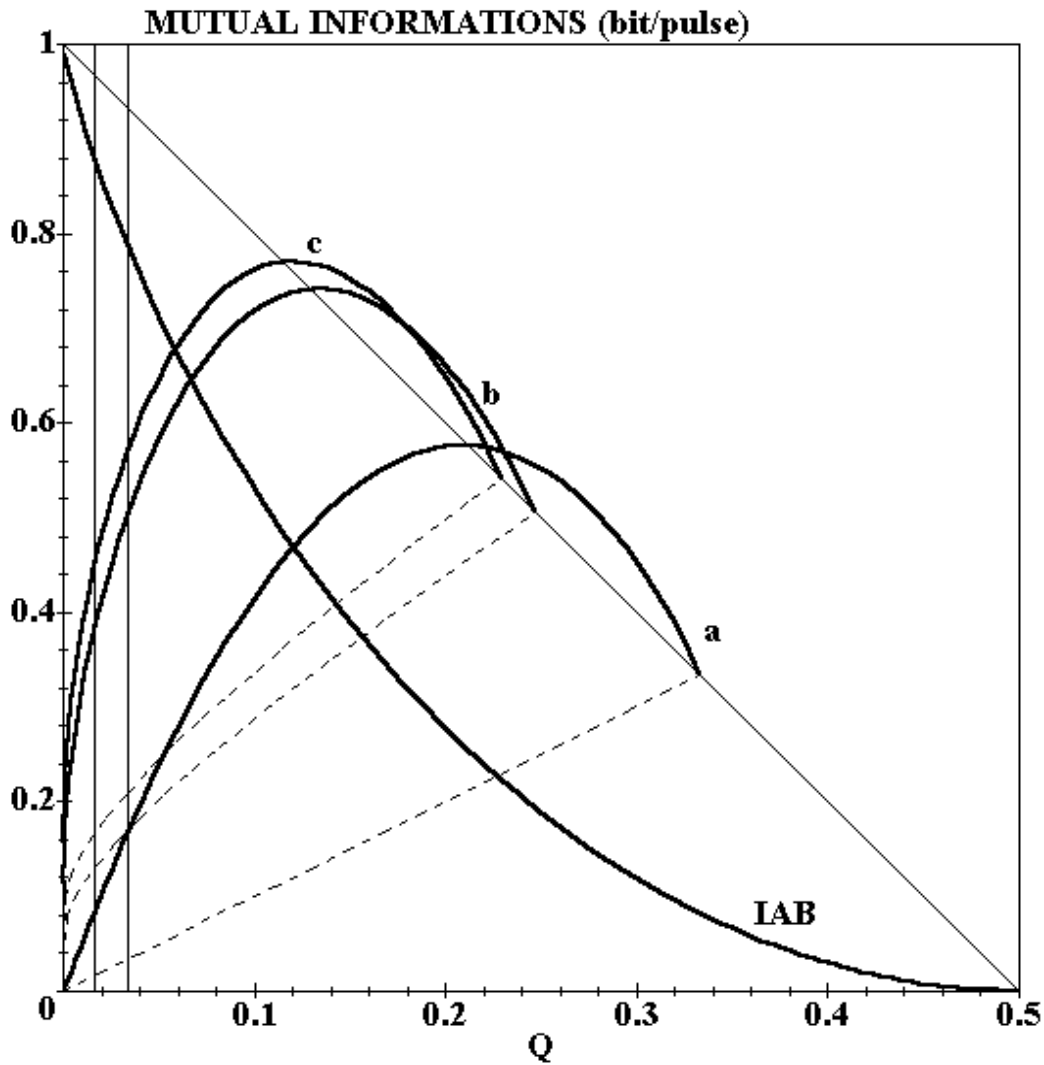

FIG. 6 : Mutual informations as a function of the QBER (Q) in the case of the attack with intrication (solid line). In each case, IAE is calculated for the ideal case and for the two experimental cases corresponding to the relative coherence $_{loss}$ $\Delta$ : 0, 0.061 and 0.086 (a, b and c respectively). They are compared with the intercept-resend attacks calculated with the same parameter (dashed line) and considering that Eve intercepts a fraction of the pulses ranging from 0 to 1. The thin line gives the maximum information that can be obtained by Eve when she intercepts all pulses sent by Alice. For a given value $_{of}$ $\Delta$, IAE is an increasing value of Q. The vertical lines corresponds to the 1.62 % and 3.3 % experimentally measured QBER values. Secure key transmission is possible when IAB is greater than **IAE.**

## TABLE CAPTION

Table I : Maximum value of the QBER in the case of the two time-slots attack, of the maximum coherence attack and of the attack with entanglement in the case of the improved protocol. They are calculated with statistical noise taken into account ( $\Delta = 0.086$ ), with no statistical noise ( $\Delta = 0.061$ ) and in the case of a perfect system ( $\Delta = 0$ ).

Table II : Information advantage of Bob on Alice corresponding to the experimental QBER of 3.3 %. They are calculated in the case of the two time-slots attack, of the maximum coherence attack and of the attack with entanglement in the case of the improved protocol. Three values of $\Delta$ are considered : with statistical noise taken into account ( $\Delta = 0.086$ ), with no statistical noise ( $\Delta = 0.061$ ) and in the case of a perfect system ( $\Delta = 0$ ).

Table III : Information advantage of Bob on Alice corresponding to the experimental QBER of 1.62 %. They are calculated in the case of the two time-slots attack, of the maximum coherence attack and of the attack with entanglement in the case of the improved protocol. Three values of $\Delta$ are considered : with statistical noise taken into account ( $\Delta = 0.086$ ), with no statistical noise ( $\Delta = 0.061$ ) and in the case of a perfect system ( $\Delta = 0$ ).

| Maximum allowed QBER in % | With statistical noise | Without statistical noise | Ideal set-up |
|---|---|---|---|
| Two time-slots attack | 9.7 | 11 | 17 |
| Maximum coherence attack | 4.6 | 5.0 | 5.8 |
| Attack with intrication (improved protocol) | 5.8 | 6.5 | 12 |

Table I : Maximum value of the QBER in the case of the two time-slots attack, of the maximum coherence attack and of the attack with entanglement in the case of the improved protocol. They are calculated with statistical noise taken into account ( $\Delta = 0.086$ ), with no statistical noise ( $\Delta = 0.061$ ) and in the case of a perfect system **( $\Delta = 0$ ).**

| Information advantage of Bob on Eve (bits / pulse) | With statistical noise | Without statistical noise | Ideal set-up |
|---|---|---|---|
| Two time-slots attack | 0.49 | 0.54 | 0.72 |
| Maximum coherence attack | 0.22 | 0.27 | 0.41 |
| Attack with intrication (improved protocol) | 0.22 | 0.29 | 063 |

Table II : Information advantage of Bob on Alice corresponding to the experimental QBER of 3.3 %. They are calculated in the case of the two time-slots attack, of the maximum coherence attack and of the attack with entanglement in the case of the improved protocol. Three values of $\Delta$ are considered : with statistical noise taken into account ( $\Delta = 0.086$ ), with no statistical noise ( $\Delta = 0.061$ ) and in the case of a perfect system ( $\Delta = 0$ ).

| Information advantage of Bob on Eve (bits / pulse) | With statistical noise | Without statistical noise | Ideal set-up |
|---|---|---|---|
| Two time-slots attack | 0.66 | 0.70 | 0.83 |
| Maximum coherence attack | 0.52 | 0.57 | 0.69 |
| Attack with intrication (improved protocol) | 0.43 | 0.50 | 0.80 |

Table III : Information advantage of Bob on Alice corresponding to the experimental QBER of 1.62 %. They are calculated in the case of the two time-slots attack, of the maximum coherence attack and of the attack with entanglement in the case of the improved protocol. Three values of $\Delta$ are considered : with statistical noise taken into account ($\Delta = 0.086$), with no statistical noise ($\Delta = 0.061$) and in the case of a perfect system ($\Delta = 0$).